%% file: main.tex
\documentclass[acmtog]{acmart}

\usepackage{comment}
\usepackage{algorithm}
\usepackage{algpseudocode}
\usepackage{amsmath}

\acmJournal{TOG}

\acmSubmissionID{673}

\begin{document}

\title{Animating Childlike Drawings with 2.5D Character Rigs}

\author{Harrison Jesse Smith}
\authornote{Corresponding author.}
\affiliation{%
  \institution{Meta}
  \city{Sausalito}
  \state{CA}
  \country{USA}
}
\email{hjessmith@gmail.com}

\author{Nicky He}
\affiliation{%
  \institution{Meta}
  \city{Sausalito}
  \state{CA}
  \country{USA}
}
\email{nicky.hsj@gmail.com}

\author{Yuting Ye}
\affiliation{%
  \institution{Meta}
  \city{Redmond}
  \state{WA}
  \country{USA}
}
\email{yuting.ye@gmail.com}

\renewcommand{\shortauthors}{Smith et al.}

\newcommand{\yuting}[1]{{\color{magenta} [Yuting: {#1}]}}
\newcommand{\jesse}[1]{{\color{blue} [Jesse: {#1}]}}
\newcommand{\nicky}[1]{{\color{green} [Nicky: {#1}]}}
\newcommand{\strike}[1]{{\color{red} [{#1}]}}

\begin{abstract}

Drawing is a fun and intuitive way to create a character, accessible even to small children. However, animating 2D figure drawings is a much more challenging task, requiring specialized tools and skills. Bringing 2D figures to 3D so they can be animated and consumed in immersive media poses an even greater challenge. Moreover, it is desirable to preserve the unique style and identity of the figure when it is being animated and viewed from different perspectives. In this work, we present an approachable and easy-to-create 2.5D character model and retargeting technique that can apply complex 3D skeletal motion, including rotation within the transverse plane, onto a single childlike figure drawing in a style-preserving manner in realtime. Because our solution is view-dependent, the resulting character is well-suited for animation in both 2D and 3D contexts. We also present a novel annotation study motivating our system design decisions and a pair of user studies validating the usefulness and appeal of our solution. We in addition showcase the generality of our system in a range of 2D and 3D applications.

\end{abstract}


\begin{teaserfigure}
  \centering
  \includegraphics[width=0.95\textwidth]{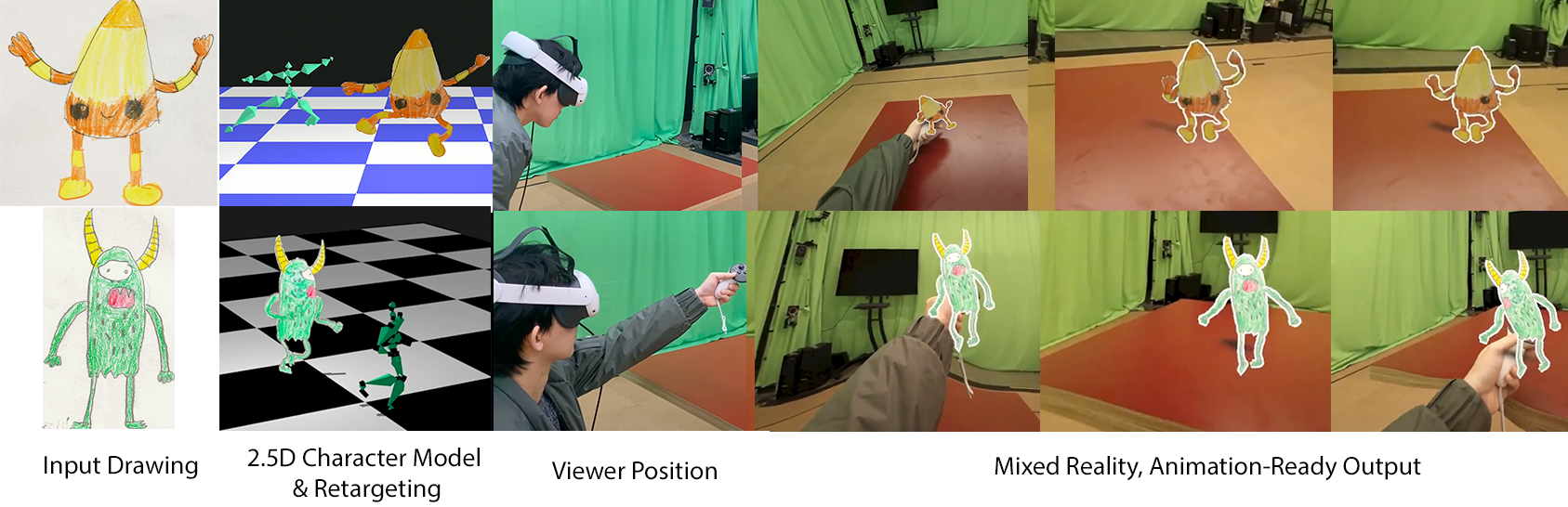}
  \caption{We present an animation system that turns a single childlike figure drawing into a 2.5D character model. It can be animated using any 3D skeletal motion and viewed from different perspectives. Our real-time and view-dependent motion retargeting algorithm makes it suitable for mixed-reality applications.}
  \Description{Banner Caption}
  \label{fig:teaser}
\end{teaserfigure}

\maketitle

\input{Introduction}

\input{RelatedWork}

\input{Defining_childlike_style}

\input{Method}

\input{Evaluation}

\input{Conclusion}

\bibliographystyle{ACM-Reference-Format}
\bibliography{Ref}

\input{FigurePage}

\end{document}

%% file: Introduction.tex
\section{Introduction}

Drawing is a fun and easy activity to create imaginary characters, especially embraced by children as a natural form of creative expression. It is accessible to virtually everyone, whereas very few possess the skills and tools required to produce appealing 3D characters. On the other hand, childlike figure drawings exhibit a unique and delightful appeal. When combined with a repertoire of interesting motions, they could be used to inject highly personalized animated characters into user generated stories, games, and interactive experiences. With recent advancements in spatial computing technology and devices, it is no longer a distant dream to embody the monster you create and smash some zombies, or dance with your own imaginary friend in the living room. That is, if we can enable everyone to easily bring their own hand-drawn characters to the 3D world and animate them in 3D.

In practice, preparing a character model to be animation-ready, or "rigging", requires highly specialized technical skills, years of training and experiences, as well as access to expensive professional software. In the case of 2D characters, their rig usually consists of many variations of the same character under different poses and views, requiring laborious manual labor to create and difficult to maintain style consistency. Generative AI models recently become an attractive and even viable option to automatically create 2D animations from an image~\cite{shi2023zero123}. However, precisely controlling the motion or preserving the character identity remain an open challenge. Moreover, in comparison to portrait photos or professional artwork, childlike drawings are extremely scarce in the training data~\cite{SmithHodgins}. It is therefore unreasonable to expect these models to perform well on animating childlike drawings. 


Even if we set aside the technical challenges, it is not obvious how to apply 3D motion to childlike drawings without significantly impacting their unique style. The crux of the problem stems from two characteristics. First, childlike drawings are \textit{view-dependent} representations, made to be observed from an angle close to the drawing plane's normal vector. Second, childlike drawings utilize an \textit{intellectually realistic}~\cite{luquet2001children} denotation system, as opposed to the more familiar, visually realistic system used in photographs and 3D renderings. This means that, instead of attempting to depict the subject faithfully from a specific point of view, the drawer instead chooses to include details which are, in accordance with their internal model of the subject, important to their identity. Therefore, creating a 3D model from the figure drawing and manipulating it to create visually realistic images would fundamentally alter the style.

Inspired by the desire for an easily accessible and efficient character creation system, and the quirky style of childlike figure drawings, we present an integrated view-dependent 2.5D character model and novel motion retargeting technique, specifically designed for childlike drawings. The system can retarget complex 3D motions, including transversal rotations, onto the drawn figures while preserving both the style and identity of the character, as well as being recognized as the input motion. Our system is informed by insights from both the field of children's art analysis, and by a novel annotation study examining how observers infer figure orientations.

Our system generates an animation-ready 2.5D character model from a single drawing and a set of high-level semantic annotations provided by the user. At run time, the character model can be automatically animated and rendered from a given 3D skeletal motion and camera positions at every frame. It is fast, requires no future information, and is well-suited for real-time applications. Any input which can convert to 3D skeletal poses (videos and pose estimation~\cite{li2021hybrik}, text-to-motion models~\cite{jiang2023motiongpt}, VR body tracking~\cite{10.1145/3550469.3555411, jiang22avatarposer}) can be used to drive the characters. The resulting animations are suitable for both 2D and 3D applications and, because the representation is view-dependent, it can be used to bring these figures into mixed reality settings (Figure~\ref{fig:teaser}). We believe this is the first system to combine 3D motion, 2D childlike drawings, and a view-dependent formulation with the goal of allowing novices to easily create and animate their own characters. 

We validate our design in several ways. We demonstrate the recognizability of the input 3D motion on the character with a perceptual user study; we provide several ablation visualizations justifying our design choices; and we compare our results with several existing methods. We also developed a wide range of applications to showcase the value and generality of our system.

Our contribution is two-fold: 
\begin{itemize}
    \item A first annotation study examining, in fine detail, how observers infer character axial rotation from childlike figure drawings. 
    \item An integrated view-dependent 2.5D character model and novel retargeting technique to apply user-generated 3D motions onto user-generated childlike figure drawings in a style-preserving manner.
\end{itemize}

%% file: RelatedWork.tex
\section{Related Work}

\subsection{View-Dependent Control}
View-dependent representations have a long history in computer graphics.
In his pioneering work, Rademacher proposed interpolating between \textit{key viewpoints} and associated \textit{key deformations} to manipulate 3D models~\cite{rademacher1999view}.
Other researchers have extended the idea to create 3D animation systems~\cite{10.1111:j.1467-8659.2004.00772.x}, streamline the modeling process~\cite{DBLP:journals/corr/abs-2103-15472}, and integrate physical simulation\cite{koyama2013view}.
Of particular note, Rivers et al.~\cite{rivers25Dcartoonmodels} introduced \textit{2.5D Cartoon Models}, a combination of planar meshes transformed, based upon view angle, so as to appears three dimensional.
Our work draws upon these works but is, to our knowledge, the first work to attempt to use view-dependent techniques to retarget 3D motion onto 2D characters.   

\subsection{Animation from 2D Images}

Many researchers have proposed different methods for creating animations from 2D images. Hornung et al.~\cite{Hornung2007anim2Dpicmotion} presented a method to deform a character from a photograph given user-provided joint annotations.
\textit{Toonsynth}~\cite{Dvoroznak18-SIG} and \textit{Neural Puppet}~\cite{poursaeed2020neural} both present methods to create new images of hand-drawn characters from examples.
Other researchers have proposed methods of obtaining 3D geometry from 2D sketches~\cite{igarashi2006teddy, Dvoroznak20-SA} and images~\cite{ArtiSketch,weng2019photo}.
A number of works have specifically focused on childlike drawings.
Lingens et al.~\cite{lingens2020towards} proposed an evolutionary algorithm for animating children's drawings. 
\textit{MagicToon}~\cite{feng2017magictoon} creates a 3D model from childlike drawings for AR applications.
Similar to our work, Smith et al.~\cite{SmithHodgins} proposed a method for animating childlike drawings using 3D skeletal motion. 
However, the resulting animations are only suitable for use in 2D applications and the type of motions it supports are limited.

Unlike these previous works, our solution can be used in 3D contexts but does not create a 3D model. We instead relying upon a view-dependent formulation of the animated character.

%% file: Defining_childlike_style.tex
\section{Orientation Depiction in Childlike Drawings}
Because our goal is to manipulate the figure in a style-preserving manner, we provide relevant insights from the fields of child psychology and art analysis about the factors which give rise to childlike drawing style. 
We also present the results of a novel annotation study exploring how observers infer the orientation of childlike drawing figures.

\subsection{How Children Draw Figures}

By childlike drawing, we refer to the quirky, representational style of drawing that tends to appear between ages three and seven~\cite{lowenfeld1975creative}. 
Though it falls out of use shortly thereafter, this style of drawing remains available to people of all ages, including adults who `cannot draw.'
This type of drawing is characterized by `intellectual realism,' a denotation system distinct from the 'visual realism' used in photographs and most 3D graphics scenes~\cite{luquet2001children}.
Such drawings occur when the drawer leverages an alternative definition of `realistic:' rather than drawing a subject as it appears when viewed from a specific vantage, they instead aims to include details important to their internal model of the subject.
Said differently, when intellectual realism is used, be it by a child or Picasso~\cite{picasso1937}, the aim is to create a recognizable representation of the subject- not necessarily a visually accurate one.

Others have argued that a key to defining childlike drawing style is understanding that the fundamental primitive used is the \textit{region}, a 2D area (e.g. a dot, single line, outlined area, or scribble cluster) whose attributes are perceptually similar to those of the subject it represents~\cite{willats2006making}. Childlike drawings can therefore be seen as a collection of regions arranged so as to convey the \textit{idea} or \textit{important details} of something in a non-visually realistic manner. 

This influences figure drawings in several ways.
First, foreshortening is rare, as long regions, such as arms, are less recognizable when foreshortened.~\cite{willats1992representation,piaget1956}.
Second, different parts of the figure are drawn so their most recognizable traits are apparent; this often results in \textit{twisted perspective}: different parts of the figure drawn as though viewed from different angles (e.g. feet in Figure~\ref{fig:part_trait_example}.a). 
Third, the majority of figures are drawn in a forward-facing 'canonical view'~\cite{goodnow1977children,cox2014drawings}, but may be drawn in profile if their orientation is important for some reason~\cite{cox1993children}.
One study found that, when drawing figures in non-forward-facing views, they omitted facial features when drawing the back side. For side views, children manipulated facial features, head contouring, obscured limbs behind each other, and pointed feet to the same side.


\subsection{How Observers Infer Figure Orientations}
\label{sec:orientation}

While it is useful to understand why children draw as they do, our aim is to modify existing figures such that the 3D pose is recognizable to an observer. Therefore, it is important to understand how observers interpret figure orientation: whether the character's forward vector extends to the left, the right, or is centered (i.e. extends directly out of the drawing). We therefore conducted an annotation study to answer the following questions:
\begin{itemize}
  \item Q1 - To what degree do observers agree on the orientation of childlike figure drawings?
  \item Q2 - What body parts and part traits do observers use when inferring orientation?
  \item Q3 - How frequently is twisted perspective present in these drawings?
\end{itemize}

\begin{figure*}[ht]
\centering
\includegraphics[width=0.9\textwidth]{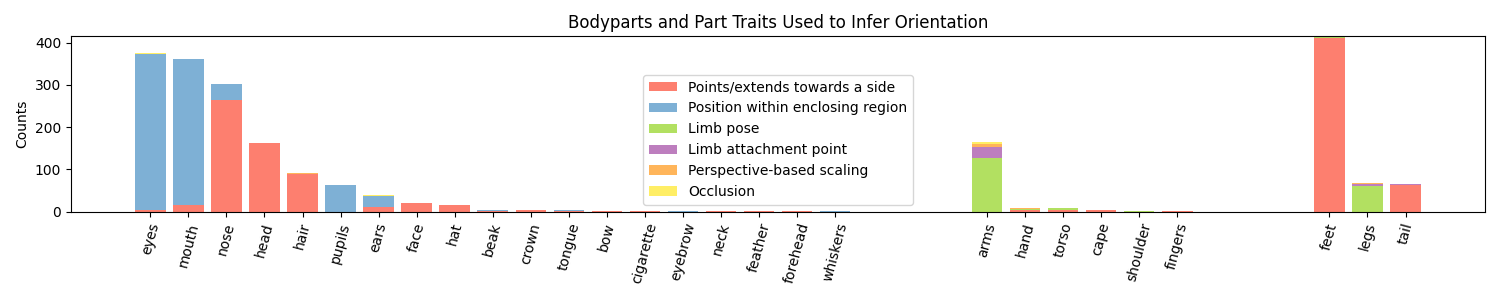}
\caption{Frequency with which annotators references specific body parts and specified part traits when determining figure orientation.}
\label{fig:part_count_orientation}
\end{figure*}

\begin{table}
  \centering
  \begin{tabular}{|c|c|}
    \hline
    Per Figure Segment Orientations & Count \\
    \hline
    Center Only & 108 (27\%)\\
    \hline
    Left Only & 4 (1\%)\\
    \hline
    Right Only & 5 (1\%)\\
    \hline
    Center, Right & 137 (34\%)\\
    \hline
    Center, Left & 124 (31\%)\\
    \hline
    Right, Left & 0 (0\%)\\
    \hline
    Right, Center, Left & 21 (5\%)\\
    \hline
  \end{tabular}
  \caption{Counts of how frequently different combinations of orientations were present in the same figure.}
  \label{table:twisted_perpspective_frequency}
\end{table}

Please refer to Appendix 1 for methodology and results tables.
Annotator agreement was substantial, indicating that observers reliably come to similar conclusions when assessing figure drawing orientations.
Figure \ref{fig:part_count_orientation} shows the frequency with which annotators used different body parts to infer orientation, along with the proportion of the times that specific part traits were mentioned.
Unsurprisingly, facial features and feet provided the majority of the orientations cues. 
There is a long tail of different body parts and accessories that could be used, depending upon what the drawer chose to include on the character.
Interestingly, only a few part traits were used the vast majority of the time.
Some parts were translated around an enclosing region, such as eyes within the face.
Other parts, such as noses, often pointed towards one direction or another.
At times the contour of a region, such as the head, extended further towards one side, suggesting orientation.
When arms and legs were used to infer orientation, they were posed such as might be seen when viewed from a particular direction. For examples of each of these, see Figure~\ref{fig:part_trait_example}.

A chart showing the frequency of twisted perspective can be seen in Table~\ref{table:twisted_perpspective_frequency}.
As expected, twisted perspective was seen frequently: 71\% of figure were composed of segments with multiple orientations.
Interestingly, at least one segment was perceived as forward facing 92\% of the time, and left-facing and right-facing segments were mixed together only 5\% of the time.

Taken together, these results suggest the following for system design:
first, it's possible to manipulate existing figure drawings to reliably influence their perceived orientation.
Second, this can be done by applying a relatively small number of manipulation types to different body parts.
Third, while it is common and acceptable to mix center-facing and consistent side-facing cues, mixing left-facing and right-facing cues should be avoided.

%% file: Method.tex
\section{Method}
Our method takes a single childlike drawing as input. The drawing should contain an upright human-like figure, and the figure must be contiguous and have no overlapping (occluded) body parts. But it can be in any pose or facing direction. 
We first process the drawing to obtaining a set of high-level annotations such as joints and body parts (Section~\ref{sec:annotation}). We then build a 2.5D model of the figure from these annotations specific for view-dependent rendering (Section~\ref{sec:model}). At runtime, our retargeting algorithm takes a viewing angle and a 3D skeletal motion, and retargets the 3D motion on the 2.5D model in a view-dependent way (Section~\ref{sec:retarget}).



\subsection{Annotations}
\label{sec:annotation}
The input drawing first goes through the annotation process as in Smith et al.~\cite{SmithHodgins} to automatically obtain 15 joint keypoints and a foreground mask of the figure, optionally with user assistance.

Next, we ask users to annotate regions of the figure with two types of orientation cues: \textit{silhouetted orientation cues} and \textit{internal orientation cues}. Starting from the figure mask as the root, these annotated regions form a hierarchy representing the figure's structure.
These two types of cues are identified based on the findings from our annotation study(Section~\ref{sec:orientation}) and the annotations inform how we manipulate the figure to depict the desired orientation from a given view.

\begin{figure}[ht]
\centering
\includegraphics[width=0.4\textwidth]{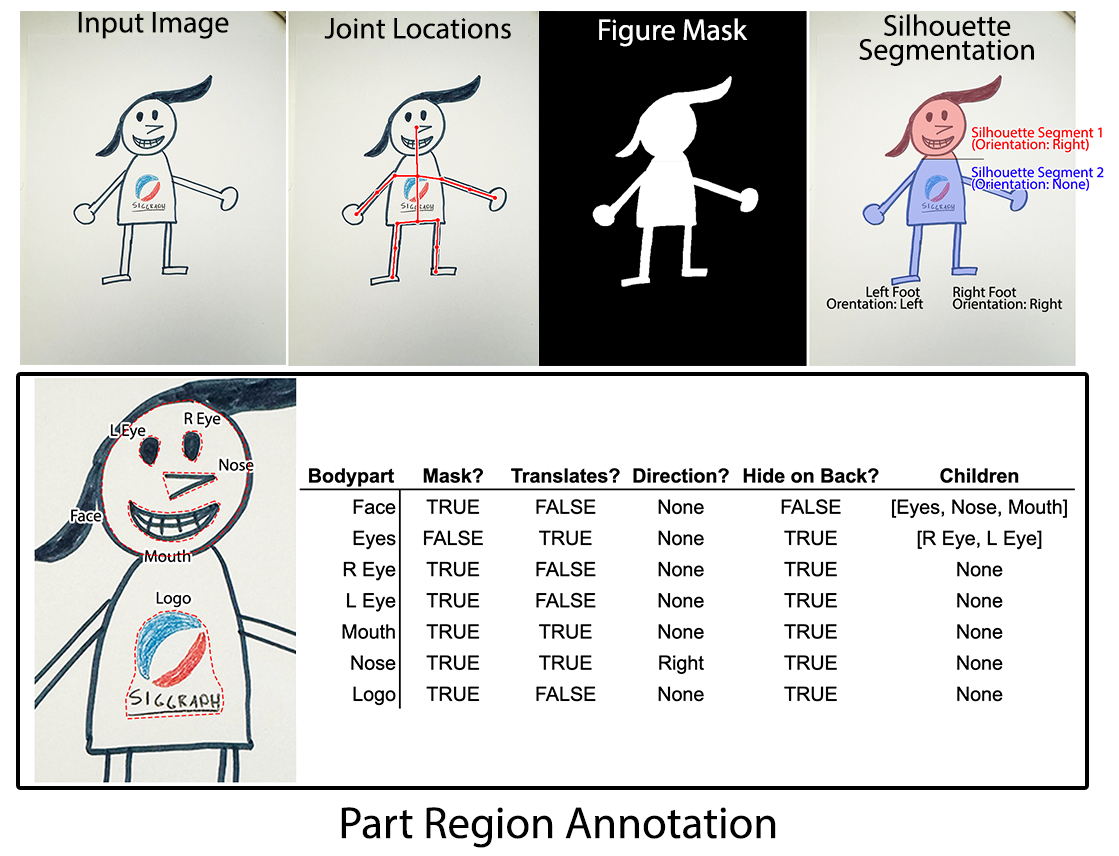}
\caption{Example of the different figure annotations and part-level attributes necessary to create the 2.5D character rigs.}
\label{fig:example_annotations}
\end{figure}

\textit{Silhouetted orientation cues} are orientation-suggesting parts of the figure visible from its outline, such as the hair in Figure~\ref{fig:example_annotations}. When applicable, users split the figure mask vertically into multiple \textit{silhouette segments}, for example into hair, head, and torso, and label each as in one of the three orientations: left, right, or none.
The feet are not separated into their own silhouette segments; rather, users indicate whether each foot is present and, if so, whether they have a left or right orientation.

\textit{Internal orientation cues} are parts of the figure that are frequently used to infer orientation but are not visible upon its outline, such as the nose, mouth, and eyes in Figure~\ref{fig:example_annotations}. A user annotates each existing cue as a \textit{part region}, and specifies five attributes on it as informed by the findings shown in Figure~\ref{fig:part_trait_example}.
\begin{itemize}
\item \textit{Mask} specifies which pixels belong in this part region.
\item \textit{Translate} indicates whether this region should translate as orientation changes, and if so whether the movement should be smooth or discrete (i.e. flipbook-like). 
\item \textit{Direction} indicates the direction of the part as left, right, or none. If the part is orienting either left or right, it may be mirrored horizontally as orientation changes. 
\item \textit{Enclosed} indicates if the part should be fully enclosed within the parent. If so, it will be hidden when it moves outside the parent region.
\item \textit{Hide on Back} indicates if the part should be hidden when the 'back' of the figure is present. 
\end{itemize}
Part regions can form a hierarchy, where a parent could be a silhouette segment or denotes a union of part regions without its own mask. 

\subsection{2.5D Character Model}
\label{sec:model}
\begin{figure}[ht]
\centering
\includegraphics[width=0.4\textwidth]{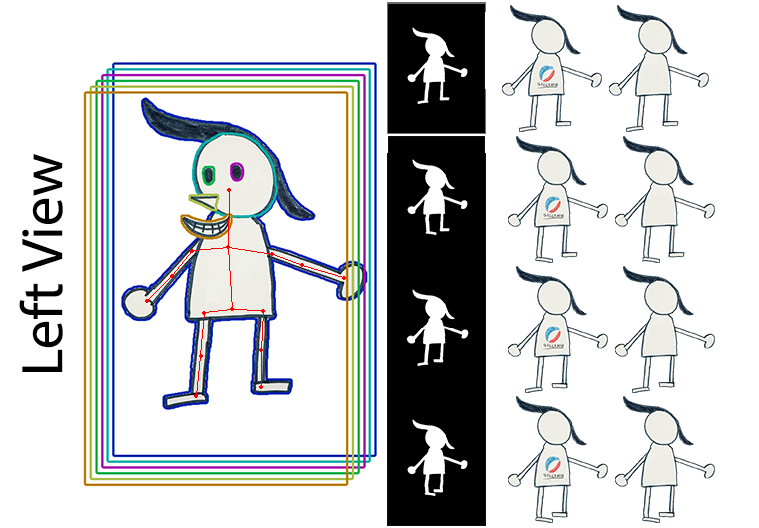}
\caption{ The part regions, masks, and textures comprising the model's left view. Each part region can be thought of as existing in a different layer (color coded here). The right view is similarly constituted, but left-facing parts, such as hair and nose, are horizontally flipped.}
\label{fig:25D_model}
\end{figure}

We construct a 2.5D character model of the figure from the annotated orientation cues. This 2.5D model is analogous to a character rig. It consists of a left-facing version of the figure and a right-facing version, termed \textit{left view} and \textit{right view} (Figure~\ref{fig:25D_model}). If the figure is drawn completely front-facing, or in other words no orientation cues are found, the two views are effectively identical. A view consists of a set of textured meshes that depict the designated character orientation with variations. It has a base mesh with two textures that indicate the front or the back of the character. The base mesh can have up to four versions of different combinations of feet orientations, depending on the foot orientation annotation result. In total, a view has up to eight unique textured meshes. In addition to the textured based meshes, each annotated part region within a view can be transformed when seen from the left or the right, as defined by a pair of \textit{keyview-transforms}. Below, we will explain how the left view is constructed as an example. The right view is created similarly.

\paragraph{Base mask}
We horizontally mirror all silhouette segments annotated as "right-facing", including their entire hierarchy. Keypoints, part regions, and child segments within the hierarchy are all mirrored as a result. Remaining part regions not belonging to any silhouette segments but are annotated as "right-facing" are also mirrored. At the end of this process, the left view will contain only "none" and "left-facing" silhouette segments and part regions (see Table~\ref{table:twisted_perpspective_frequency}).

\paragraph{Textures} When part regions move around, they leave behind holes in the texture. We therefore use image inpainting tools \cite{} to create a front texture and a back texture. The front texture is inpainted after removing all part regions that can translate. The back texture is inpainted by additionally removing static part regions that only show on the front (eg. buttons on a shirt).

\paragraph{Foot orientations} If a foot is annotated as either left-facing or right-facing, we mirror it in place to create the other version; this results in up two four versions of the base mask when both feet have orientation. We define the foot region as starting midway between the ankle and knee keypoints, and both the mask and the corresponding textures are mirrored and stitched back to the lower leg (see Figure~\ref{fig:25D_model}).

\paragraph{Mesh Generation} From each mask, we extract a contour using the marching square algorithm~\cite{lorensen1998marching} generate a mesh with Triangle's implementation of constrained Delaunay triangulation~\cite{shewchuk96b}, and texture it with front and back textures.

\paragraph{Keyview-transform} A translating part region has two key positions when seen from the left or the right, respectively. We define a $3x3$ transformation for each viewing angle to specify how the \textit{anchor point} of a part region should be transformed. This results in two (key view, transform) pairs for each part region. The definition of a anchor point depends on the part region's mask and orientation. If a part region is without left-right orientation, the anchor is simply the centroid of the mask. If it is left-facing, the anchor point is instead positioned on the right side of the region's bounding box; right-facing parts are handled inversely. If the part has no mask, its centroid is the average of its children's centroids.

An example 2.5D character model is shown in Figure~\ref{fig:25D_model}. It consists of two \textit{views}, each containing a single set of joint keypoints and a maximum of four masks and two textures per mask. Each part region lies upon its own layer and contains two keyview-transform pairs.



\subsection{View-dependent Animation Retargeting}
\label{sec:retarget}
Our retargeting algorithm animates the 2.5D character model on a vertical 2D plane given a 3D skeletal motion and a camera view. It consists of three steps: defining a view-dependent 2D character plane (Section~\ref{sec:retarget-plane}), retargeting 3D animation as 2D character poses (Section~\ref{sec:retarget-pose}), and deforming the character mesh based on the 2D poses (Section~\ref{sec:retarget-mesh}). To mitigate major artifacts in 3D to 2D projection, we additionally introduce a novel projection plane optimization algorithm using the principle of twisted perspective in Section~\ref{ComputeRigJointLocations}. It improves the motion smoothness and recognizability of the character when used for pose retargeting.

\subsubsection{Character plane}
\label{sec:retarget-plane}
\begin{figure}[ht]
\centering
\includegraphics[width=0.4\textwidth]{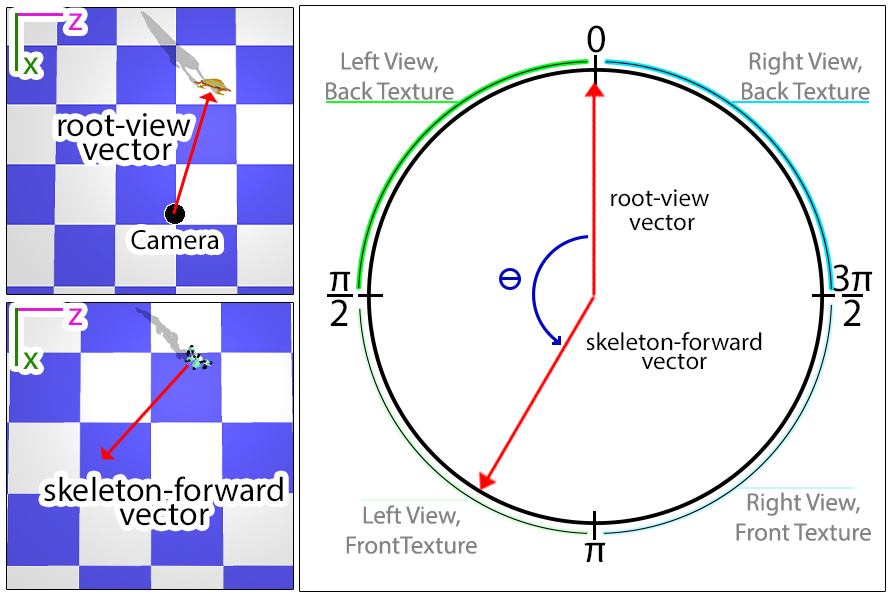}
\caption{View unit circle.  Root view vector (define as ground projection of vector from camera to character) is always at 0.Change skeleton to pose. When view angle = 0, character is looking away. pi/2 = looking left. pi = looking directly at character 3pi/2 = looking right.}
\label{fig:view_circle}
\end{figure}

The 2.5D character lives on a 2D \textit{character plane} that rotates about the vertical axis to face the camera, and translates to follow the character. We formally define the view angle $\theta$ as the angle between the \textit{root-view vector} and the \textit{skeleton-forward vector} (see Figure~\ref{fig:view_circle}). The \textit{root-view vector} is the ground projection (i.e. removing pitch and roll) of the 3D vector from the camera to the 2.5D character's root joint
. The \textit{skeleton-forward vector} is the ground projection of the normal vector of the skeleton's frontal plane, defined by the three vectors connecting the shoulder joints and the hip joint. The character plane then rotates to align its normal vector with the negative root-view vector. The position of the character plane follows the character's root joint. The character root is updated by scaling the skeleton's root velocity by the leg length ratio between the skeleton and the character.





\subsubsection{Character pose retargeting}
\label{sec:retarget-pose}

After we decide on the character plane, we need to choose between the two character views as the target for pose retargeting. We use the left view when $\theta \in [0, \pi]$, and the right view when $\theta \in [\pi, 2\pi)$. The character joints defined by the keypoints in the chosen view serves as the target 2D character.

Our retargeting follows Smith et. al. \cite{SmithHodgins}, assuming a joint mapping is available between the skeletal and the character. For a pair of corresponding joints, we project the 3D joint vector to the character plane, and rotate the corresponding character joint in 2D to align their directions. This process obeys two major constraints. Firstly, the character joints are constrained to 2D rotations in-plane. Secondly, in accordance with the style of childlike drawings, the character joints are not foreshortened but maintain their lengths.

In the case when the skeleton is back-facing the camera (i.e. $\theta \in [0, \frac{\pi}{2}], \text{or } \theta \in [\frac{3\pi}{2}, 2\pi)$), we swap the character's left limbs with the right limbs in the joint mapping to "turn $180^{\circ}$". It means the left limbs of the skeleton drive the right limbs of the character, and vice versa.

Note that the feet joints are excluded in this retargeting process. We instead determine the feet orientations based on the knee orientations (see below) to prevent them from appearing "bending backwards".



\subsubsection{Character mesh deformation}
\label{sec:retarget-mesh}
The character pose is used to deform the textured character mesh of the chosen view. We first choose the final rest mesh based on the feet orientation. The sidedness of a foot is the same as where the knee is pointing on the 2D plane. Once the final mesh is selected, it is deformed using ARAP~\cite{igarashi2005asrigidaspossible} with the posed character joint locations as control handles. 
We then use the view angle to interpolate between each part region's keyview-transform pairs and to render them with proper translation and rotation upon the deformed mesh.



Because the character exists upon a 2D plane, depth-based culling is insufficient to properly obscure intersecting body parts. Instead, we explicitly compute the render order of each mesh triangle, along with the part regions attached to those triangles. We determine render order by analyzing the input 3D skeleton's joint positions and ordering them based upon depth to their projection plane.
We then render, from furtherest to nearest, the triangles mapped to the equivalent joints within the character.



\subsubsection{Projection Plane Optimization}
\label{ComputeRigJointLocations}

When projecting the 3D skeletal pose to the character plane, we observe two major artifacts. When a 3D joint is pointing near the plane normal, a small movement can cause large motion in the corresponding 2D joint, an effect we termed \textit{flailing}. On the other hand, a large movement in the 3D joint could result in no change of the corresponding 2D joint, what we termed \textit{dampening}. This is when the 3D movement only changes the magnitude of the projected 2D vector but not its angle.


We can better understand these artifacts by looking at how the 2D joint orientation \(\alpha\)\ change as a function of the 2D vector $\mathbf{P}$, where $\mathbf{P} = [(\mathbf{I} - \mathbf{Z}\mathbf{Z}^{T})\mathbf{V}]_{xy}$ is the 2D projection of the 3D vector $\mathbf{V}$ onto the XY-plane (i.e. the projection plane). We have $\alpha = atan2(P_{y}, P_{x})$, and its Jacobian is

$$
\frac{\partial \alpha}{\partial \mathbf{P}}
=
\left[ 
    \frac{-P_y}{P_x^2 + P_y^2}, 
    \frac{P_x}{P_x^2 + P_y^2}
\right].
$$

When $\mathbf{V}$ is close to the $\mathbf{Z}$ axis, $\mathbf{P}$ becomes small, and the norm of the Jacobian becomes large. Therefore, a small change in $\mathbf{P}$ results in a large change in $\alpha$. When $\mathbf{V}$ rotates about the axis $[-P_y, P_x, 0]$ or $[P_y, -P_x, 0]$ on the XY-plane, $\mathbf{P}$ changes in the direction of $[P_x, P_y]$ (i.e. $\Delta \mathbf{P} = [P_x, P_y]r, r \in \mathcal{R}$), and we have $\frac{\partial \alpha}{\partial \mathbf{P}} \Delta \mathbf{P} = 0$.

We can avoid both issues if we choose the projection plane based on the orientation and rotation axis of joints, rather than being constrained to the character plane. Such an approach is in fact supported by the use of twisted perspective frequently seen within childlike drawings (Table~\ref{table:twisted_perpspective_frequency}), where different projection planes are applied on different body parts.

We therefore formulate a multi-objective optimization to find the optimal projection plane $\mathbf{n}$ for each limb independently, since the flailing and dampening artifacts are most severe for limbs. 

Because both the knees and elbows are hinge joints, their rotation axis is the normal of the plane defined by the upper limb $\mathbf{v}_u$ and the lower limb $\mathbf{v}_l$ of the 3D skeleton. We want to prevent the rotation axis from lying on the projection plane, and want the limbs to be pointing away from the projection normal. It means $\mathbf{n}$ should be as aligned with the rotation axis as possible. We also want to encourage the projection plane to be close to the character plane, so the twisted perspective is not abused, and the 2D pose remains recognizable as a projection of the 3D pose. Additionally, we want $\mathbf{n}$ to be close to previous frame's projection plane normal $\mathbf{v}_p$ for temporal consistency.

If we represent rotations on a unit sphere, $\mathbf{v}_l$ and $\mathbf{v}_u$ forms two points on a great circle that $\mathbf{n}$ should stay away from, and the character plane normal $\mathbf{v}_c$ and $\mathbf{v}_p$ are two attractors. The closer $\mathbf{v}_l$ and $\mathbf{v}_u$ are, the more straight the limb becomes, until the limb is perfectly straight and the two vectors are collinear. We therefore scale the influence of the great circle by the angle between the two vectors. 

Formally, we solve for $\mathbf{n}$ on the unit sphere by minimizing the following cost function:

\begin{equation}
\min_{\mathbf{n}} \left( (1 - \frac{|\mathbf{v}_u^T\mathbf{v}_l|}{\|\mathbf{v}_u\|\|\mathbf{v}_l\|}) \cdot e^{\frac{d_{xt}(\mathbf{n}, \mathbf{v}_u, \mathbf{v}_l)^2}{{2 \sigma_1^2}}} + e^{\frac{d_{gc}(\mathbf{n}, \mathbf{v_c})^2}{{2 \sigma_2^2}}} + e^{\frac{{d_{gc}(\mathbf{n}, \mathbf{v_p})^2}}{{2 \sigma_3^2}}} \right)
\end{equation}

where $d_{xt}$ is the \textit{cross-track distance} between $\mathbf{n}$ and the great circle formed by $\mathbf{v}_l$ and $\mathbf{v}_u$,
and $d_{gc}$ is the \textit{spherical distance} between two points on a sphere.
Instead of computing $\mathbf{n}$ every frame for every limb, we can first check the value of $d_{xt}(\mathbf{v}_c, \mathbf{v}_u, \mathbf{v}_l)$ for a limb. If it is above a threshold, we can comfortably use $\mathbf{v}_c$ for that limb.

%% file: Evaluation.tex
\section{Evaluation}

We evaluate our system in several ways.
First, we conduct a perceptual user study validating that 3D motion, including transversal rotation, remain recognizable when applied to our 2.5D models.
Second, we present a series of ablations demonstrating the results of the system design decisions.
Third, we compare the resulting animations to existing methods. Our results and evaluation settings are best assessed from the supplementary video.

\subsection{Motion Recognition User Study}
\label{sec:recognition}
We conducted a user study to determine whether the manipulations applied to the 2.5D character models were sufficient to recognize the original 3D motions.
See video for example stimuli and Appendix 2 for full details.
Ten different MTurk HITs were generated, each showing videos of eight motion-camera position pairs applied to a different 2.5D character model and to a 3D model.
All users within a HIT saw all 16 videos.
Results were treated as nominal (correct vs. incorrect) and paired (same motion applied to 2.5D and 3D character). We therefore use the McNemar test for significance.
Results are shown in Figure~\ref{fig:user_study_1}. In nine of ten cases, there was no significant differences, indicating users were able recognize the original 3D motion at similar rates when viewed upon either a 3D or a 2.5D model.

These results show that, in general, the techniques employed by the 2.5D model are sufficient to preserve the identity of the original 3D motion, even in the absence of root translation.
However, for drawing 7, correct responses were low for two specific stimuli.
In both cases, the limb mapping switched during the clip (see Section~\ref{sec:retarget-pose}).
This character was drawn holding a carrot in one hand; the carrot did not switch hands during the limb mapping switch.
The contradiction may explain the significant difference, and indicates that held object and other limb asymmetries may need to be handled differently.



\subsection{Style Preference User Study}
\label{sec:style}
To validate the appeal of our method, we conducted an exploratory second user study asking viewers to rank a basic animated scene among a 2D, 2.5D, and 3D character representations.
The 3D models were hand-crafted by a professional 3D artist.
45 MTurk workers were hired and asked to choose the most and least style-preserving, identity-preserving, visually appealing, scene-cohesive, and personally preferred representation. Stimuli are shown in the supplementary video. Additional study details are given in the Appendix 3. 
For two characters,\textit{ Candy Corn} and \textit{Stick Girl}, 3D and 2.5D methods were preferred at similar levels.
For \textit{Robot}, 3D was preferred. 
2D was always least preferred among all characters.
This suggests that our 2.5D model is a valuable stylistic alternative to 3D, in addition to being much easier to create.
The justifications users provided for their opinions were informative. While some people preferred the professional polished look of 3D, others felt it fundamentally altered the character in a way the 2.5D method did not.

\subsection{Ablation}
To justify our design decisions, we provided several ablations.

\paragraph{View-dependence.}
Childlike drawings are drawn in a view-dependent manner. To preserver their unique style, it is necessary to manipulate them in a similarly view-dependent way. To demonstrate its importance, we provided side-by-side comparisons of view-dependent and non-view-dependent animation in the video. In addition to changing the style of the figure, the motion of the limbs is less clear and the character disappears completely from certain viewing angles without our algorithm.

\paragraph{Twisted Perspective Retargeting}
We provide examples of a retargeted pose sequence with single versus multiple projection planes (see video).
When the axis of the skeleton's left lower arm approaches the root-view vector, the arm flails suddenly (Figure~\ref{fig:twisted_perspective} top right). 
This is corrected (Figure~\ref{fig:twisted_perspective} bottom right) by automatically selecting a more suitable projection plane (Section~\ref{ComputeRigJointLocations}).

\paragraph{Limb Swapping}
When a skeleton's 3D joint positions are projected onto a 2D plane, the relative position of the limb joints depends upon the skeleton's transversal rotation. 
When facing the camera, the skeleton's left limbs appear to the camera's right side; when facing away, this is reversed. 
Flipping the limb mapping (Section~\ref{sec:retarget-pose}) therefore serves to keep the pose recognizable (Figure~\ref{fig:twisted_perspective}).

\subsection{Comparisons}
In the accompanying video, we provided comparisons against Smith et al.~\cite{SmithHodgins} and \textit{Photo Wake-Up}~\cite{weng2019photo}.
Smith et al.'s output is inherently 2D and does not account for incongruous character orientation cues, such as foot direction. 
Photo Wake-Up creates a textured 3D object, which may fundamentally alter the style of abstract characters, such as those in Picasso paintings.

In Figure~\ref{fig:genai}, we compare novel views of a character generated with our approach to those generated by a state-of-the-art single image to multi-view diffusion model~\cite{shi2023zero123} trained primarily on visually realistic data, revealing its 3D bias.



\section{Applications}
Our system is easily accessible to non-technical users, as it requires only a single drawing as input and a set of high-level semantic annotations that are intuitively understandable. Most drawings we tested can be annotated in under three minutes even with a duck-taped together interface. Creating the 2.5D model from the annotations then takes under ten seconds on a Macbook Pro, allowing for rapid iteration.

Once the 2.5D model is generated, it can be used in a variety of different context. For example, we can animate it with any 3D motion capture data. We can then render the retargeted character on the original drawing with a fixed camera to create a 2D animated story (Figure~\ref{fig:applications}.b), or we can render it in 3D to create a stylized animated shorts with a dynamic camera (Figure~\ref{fig:applications}.a,.d). 

Our system can be combined with any technique that generates 3D skeletal motion as output. As an example, we retargeted the human motion generated by a text-to-motion model~\cite{Qian_2023_ICCV} on a figure drawing as an inspiration for story telling (Figure~\ref{fig:applications}.c). W can also puppeteer characters with any real time body tracking algorithms, even in VR~\cite{10.1145/3550469.3555411} (Figure~\ref{fig:applications}.e), as the retargeting algorithm runs comfortably at over 30fps on an M1 Macbook Pro.



The view-dependent element of our model makes it particularly well-suited for mixed reality applications, in which the user has control over the camera position. We demonstrate this by creating a mixed reality application allowing users to view and pick up characters (Figure~\ref{fig:applications}.f).






%% file: Conclusion.tex
\section{Conclusion}

We present an integrated view-dependent 2.5D character model and novel motion retargeting technique to animated childlike figure drawings with 3D skeletal motion.
It requires only a single image and a set of high-level annotations, which are easy for a user to provide, and can be used to animate the figures in real-time.
The types of manipulations supported by our system are informed by the field of children's art analysis and by a novel annotation study we conducted.
Though simple, they are effective at keeping the motion identifiable, as indicated by the motion recognition study. 
As indicated by the style preference study, our results are appealing, style-preserving, and present a viable alternative to full 3D models. We also compare our results with alternatives, revealing limitations of existing methods in animating 2D figures in 3D. However, there is no "ground truth" for what an imaginary figure should look like, except for what is in the creator's mind. Quantitative evaluations therefore remain an important and fruitful future direction.

Our system currently requires users to provide a set of high-level semantic annotations. This task can be accelerated by latest AI models such as SAM~\cite{kirillov2023segment} and a more streamlined user interface.
However, there are certain limitations to our system that tools cannot overcome and are worth further investigation.
We require the character to be bidpedal and have no overlapping parts. 
The character must also be approximately vertical so that horizontally mirroring a region switches it from left- to right-facing.

There are also limitations to what 3D motions are applicable, either due to the nature of 2D projection, or the limitation in assuming only transversal rotations in the algorithm. For instance, motions that require three dimensions, such as pointing at the camera, won't appear as desired. Planar motions oriented towards the camera, such as someone cartwheeling towards you, may also fail.

The system produces animations that are not visually realistic, and therefore may look undesireable when used to animate visually realistic figures. An additional outcome of this design decision is that interactions with external objects won't appear realistic. The most obvious example of this is foot sliding, which can occur both as a result of using multiple projection planes for retargeting as well as changing the position of the viewing camera.


%% file: FigurePage.tex
\begin{figure*}[ht]
\centering
\includegraphics[width=0.9\textwidth]{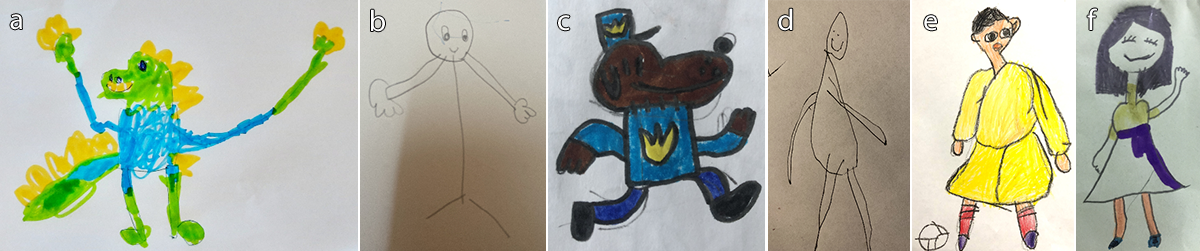}
\caption{Examples of the different part traits mentioned by annotators. The extension of the tail and nose in \textit{a} suggest (conflicting) orientations; in \textit{b} the position of the eyes within head and pupils within eye suggest it; in \textit{c} the limbs are drawn as though viewing a character facing right; \textit{d-f} show less frequently mentioned cues of limb attachment point, perspective-based scaling, and occlusion, respectively.}
\label{fig:part_trait_example}
\end{figure*}

\begin{figure*}[ht]
\centering
\includegraphics[width=0.9\textwidth]{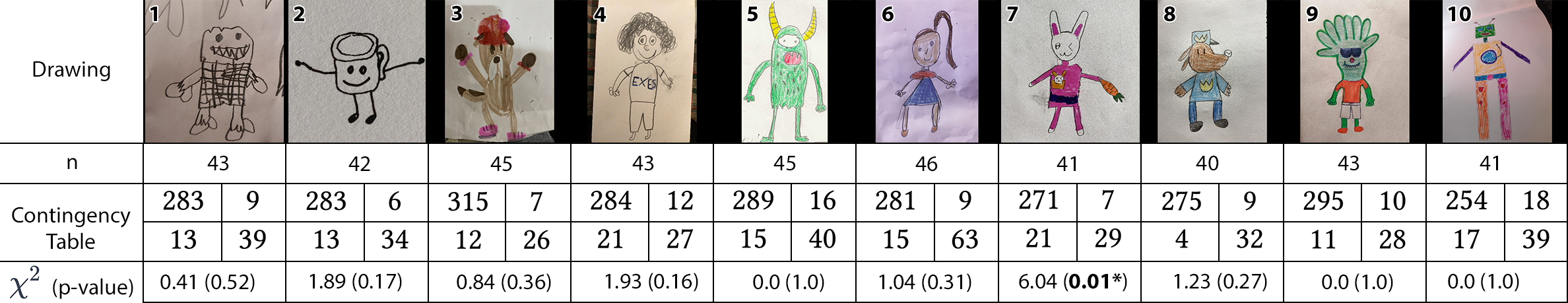}
\caption{\textit{Top row:} original drawings of the 10 characters used in the perceptual user study. 
Subsequent rows contain the number of participants in each study, response contingency table, and chi-squared test static with p-value. The upper left cell of the contingency table contains the number of times a user correctly identified the original motion upon both the 2.5D model and the 3D model; upper right contains times motion was correctly identified upon 2.5D model only; lower left contains times motion was correctly identified upon 3D model only; bottom right contains times motion was not correctly identified in either case.}
\label{fig:user_study_1}
\end{figure*}

\begin{figure*}[ht]
\centering
\includegraphics[width=0.9\textwidth]{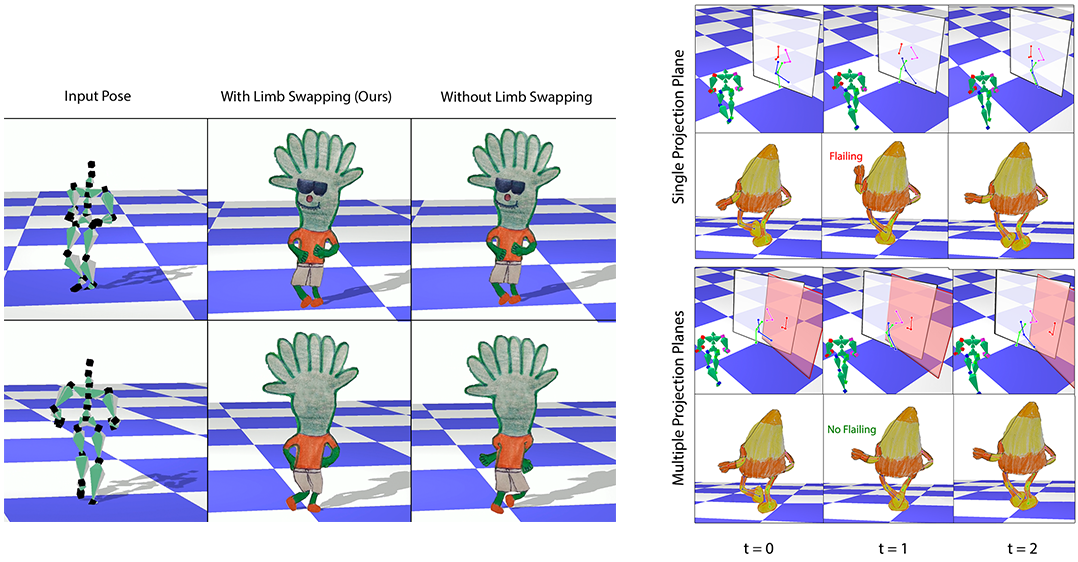}
\caption{
\textit{Left:} when the input pose faces towards the camera (top row), it's left limbs are used to drive the limbs on the \textit{drawing-right} side of the model.
When the input pose faces away from the camera, (bottom row), this mapping is flipped and the left limbs now drive the limbs on the \textit{drawing-left} side of the model (middle column).
Without this flipping, limbs appear switched and the pose is less recognizable (right column, arms.
\textit{Right:} when only the root-view projection plane  (shown in white) is used for retargeting, unexpected flailing can occur; when the axis of the left lower arm is parallel to the root-view projection plane normal vector, flailing can occur (top, middle column). 
Dynamically modifying the limb's projection plane based addresses this; in the bottom rows, the skeleton's left arm projection plane (shown in red) deviates from the root-view projection plane, preventing flailing from occurring.}
\label{fig:twisted_perspective}
\end{figure*}



\begin{figure*}[ht]
\centering
\includegraphics[width=1.0\textwidth]{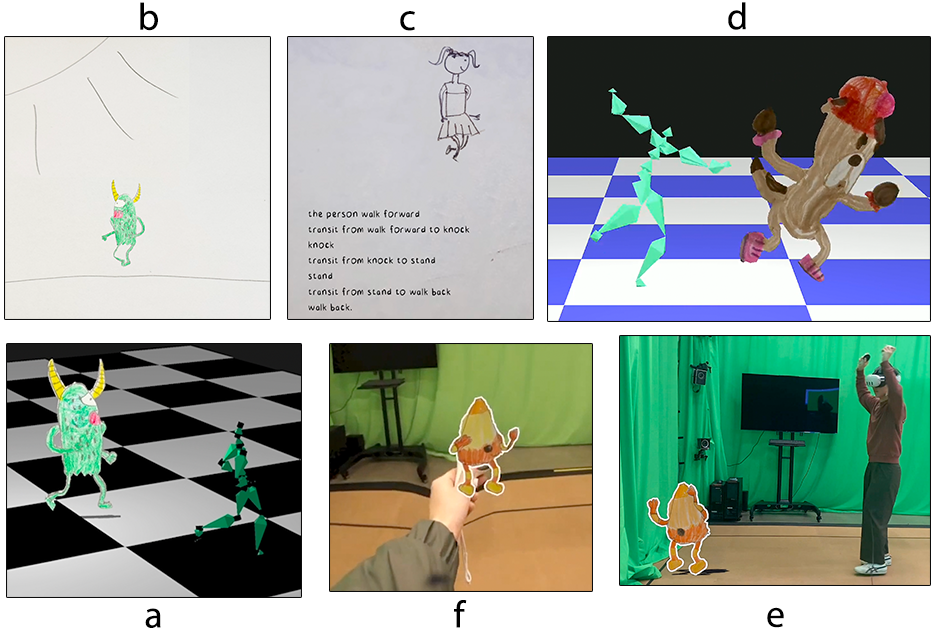}
\caption{Example applications: 3D scenes (a), 2D scenes (b), driving with text-to-motion model (c), motion capture (d), and 3-point tracking (e). Example mixed reality experience (f).}
\label{fig:applications}
\end{figure*}

\begin{figure*}[ht]
\centering
\includegraphics[width=1.0\textwidth]{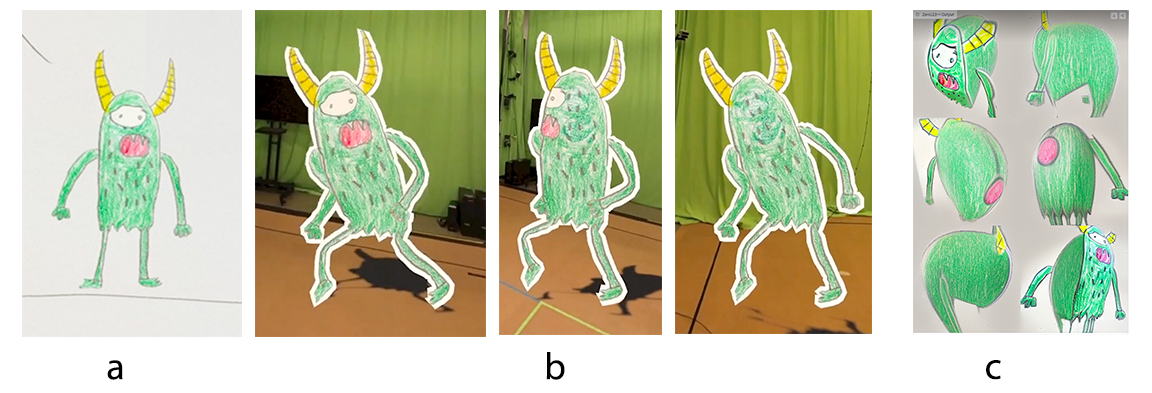}
\caption{Original drawing (a), alternate views created by our method (b), and alternate views created by multi-view diffusion model~\cite{shi2023zero123} (c).}
\label{fig:genai}
\end{figure*}